\documentclass[aps,prd,a4paper,nosuperscriptaddress,nofootinbib,showpacs,twocolumn,showkeys,amsfonts,amssymb,amsmath]{revtex4-2}
\usepackage{amssymb,latexsym}

\usepackage{color}
\usepackage{amscd}
\usepackage{times}
\usepackage{epsfig}
\usepackage{psfrag}
\usepackage{graphicx}
\usepackage{amsthm,amsmath}
\usepackage{mathrsfs}
\usepackage[colorlinks=true, allcolors=blue]{hyperref}
\usepackage{leftindex}
\usepackage[dvipsnames]{xcolor}
\usepackage{physics}
\usepackage{booktabs}
\usepackage{makecell}
\usepackage{adjustbox}

\begin{document}
\title[]{Spin contrast, finite temperature, and noise in matter-wave interferometer}

\author{Tian Zhou $^{1}$}
\affiliation{ 
$^{1}$ Van Swinderen Institute, University of Groningen, 9747 AG Groningen, The Netherlands}
\author{Ryan Rizaldy $^{1}$}
\affiliation{ 
$^{1}$ Van Swinderen Institute, University of Groningen, 9747 AG Groningen, The Netherlands}
\author{Martine Schut $^{2}$}
\affiliation{$^{2}$ center for Quantum Technologies, National University of Singapore, 3 Science Drive 2, 117543}
\author{Anupam Mazumdar $^{1}$}
\affiliation{ 
$^{1}$ Van Swinderen Institute, University of Groningen, 9747 AG Groningen, The Netherlands
}

\begin{abstract}
In this paper, we will show how finite-temperature corrections and spin-dependent/independent noise will affect the contrast in a matter-wave interferometer, especially with massive objects and large spatial superposition sizes. Typically, spin is embedded in a nanoparticle as a defect, which can be manipulated by the external magnetic field to create a macroscopic quantum superposition. These massive matter-wave interferometers are the cornerstone for many new fundamental advancements in physics; particularly, macroscopic quantum superposition can use entanglement features to, e.g., test physics beyond the Standard Model, test the equivalence principle, improve quantum sensors, and test the quantum nature of spacetime in a lab.  
We will consider a Stern-Gerlach-type apparatus to create macroscopic quantum superposition in a {\it harmonic oscillator} trap, and figure out the spin contrast loss due to linear spin-independent and spin-dependent noise in a single interferometer. We will show that spin contrast loss due to spin-independent noise does not depend on the initial thermal state of the matter wave function. However, spin contrast loss due to spin-dependent fluctuations is dependent on the initial thermal occupation of the quantum state. We will keep our discussion general as far as the noise parameters are concerned.

\end{abstract}

\maketitle
\section{introduction}

Spatial superpositions with nanoparticles, generally controlled by an embedded electronic spin defect (colour center), have a multitude of fundamental and commercial applications. 
One such example will be in the context of a spin defect embedded as a nitrogen-vacancy (NV) center in a nanodiamond~\cite{Doherty_2013}. There are numerous applications in quantum metrology to quantum sensors~\cite{Doherty_2013}. Furthermore, by creating spatial superposition, we open a new vista for testing decoherence effects in a matter-wave interferometer~\cite{bassireview}, detecting external accelerations due to gravity, such as a gravimeter~\cite{Wu:2022rdv,Wu:2024bzd,Toros:2020dbf}, external sources of electromagnetic interactions~\cite{Schut:2023tce}, and high-frequency gravitational waves~\cite{Marshman:2018upe}. 

One can use two such interferometers adjacent to each other to witness entanglement, as in the case of a C-NOT gate~\cite{Wineland:1992,Wineland:1995}. The entanglement is a bonafide quantum entity, which signifies the quantum correlation, starkly different from the classical correlation~\cite{Horodecki:2009zz}. It is well-known that entanglement between two quantum systems requires quantum interaction, or quantum mediator, which is the essence of a theorem known as local operations and classical communication (LOCC), which cannot entangle the two quantum systems. Hence, witnessing entanglement between two adjacent matter-wave interferometers proves to be even more fundamental than ever thought before. We can use it as a testing ground for physics beyond the Standard Model, such as detecting the fifth force, the possibility of extra $U(1)$ mediated interactions such as a hidden photon and an axion mediated interaction~\cite{Barker:2022mdz}, testing the quantum equivalence principle~\cite{Bose:2022czr}.

Therefore, authors of \cite{Bose:2017nin,ICTS,Marshman:2019sne,Bose:2022uxe}, see also  
\cite{Marletto:2017kzi} proposed to use an entanglement witness as a way to test the quantum nature of spacetime in a lab, see also~\cite{Danielson:2021egj,Carney_2019,Carney23_nu,Biswas:2022qto,christodoulou2023locally,christodoulou2019possibility,Rufo:2024ulr}. Furthermore, entanglement could also be witnessed for the relativistic corrections to the Coulomb potential~\cite{Toros:2024ozf}, and post-Newtonian corrections to quantum gravity~\cite{Toros:2024ozu}. Testing massive gravity~\cite{elahi2023probing}, and the quantum version of the modified theories of gravity in a lab~\cite{Vinckers:2023grv,chakraborty2023distinguishing}. Moreover, testing the quantum analogue of light bending experiment in the context of witnessing entanglement between matter and photon degrees of freedom~\cite{Biswas:2022qto} will embolden the spin-2 nature of the graviton as a mediator.

One probable way to create macroscopic quantum superpositions is by applying the Stern-Gerlach force~\cite{Margalit:2020qcy,amit2019t} to the NV spin embedded in a nanodiamond. The spin is susceptible to the external inhomogeneous magnetic field required to create the spatial superposition, see~\cite{Wan16_GM,Scala13_GM,
Bose:2017nin,Pedernales:2020nmf,Marshman:2021wyk,Marshman:2018upe,Zhou:2022epb,Zhou:2022frl,Zhou:2022jug,Zhou:2024voj,Braccini:2024fey,Zhou:2024pdl,Rizaldy:2024viw,japha2022role,japha2021unified}. Of course, any matter-wave interferometer is sensitive to external noise and fluctuations in ambient pressure, temperature, current, voltage, etc. ~\cite{Toros:2020dbf,Rijavec:2020qxd,
vandeKamp:2020rqh,Schut:2021svd,Schut:2023eux,Schut:2023hsy,Fragolino:2023agd,Schut:2024lgp}. There are phonon-induced noise~\cite{Henkel:2021wmj,Henkel:2023tqe,Xiang:2024zol}, and fluctuation in the spin degrees of freedom during the dynamics of rotation of the rigid body~\cite{Japha:2022phw,Zhou:2024pdl}, all leading to dephasing and decoherence, see~\cite{bassireview,ORI11_GM,Hornberger_2012}, and loss of contrast~\cite{Englert,Schwinger,Scully,Margalit:2020qcy}. One might expect that we might be able to cool the initial state of the center-of-mass motion~\cite{Deli__2020,Piotrowski_2023,Kamba:2023zoq,Bykov:2022xji,Perdriat:2024xiy}; nevertheless, it is important to know to what extent the initial state of the matter wave interferometer will affect the final contrast, known as the \emph{Humpty-Dumpty} problem, coined by the authors~\cite{Englert,Schwinger,Scully}. 

This paper will aim to provide an analysis of spin-independent and spin-dependent noise in the matter-wave interferometer by taking finite-temperature corrections to the initial state preparation. 
We will assume a simple harmonic oscillator potential or shifted harmonic oscillators for a spin system to analyze the spin contrast upon finishing the one-loop interferometer. 
In particular, we will show that the initial state of finite temperature has no bearing on the spin contrast of such a shifted harmonic oscillator-based matter-wave interferometer if the external noise is spin-independent. This is because the noise affects both arms of the interferometer in such a way that only a relative phase difference exists. However, random fluctuations of the relative phase $\delta\phi$ can also cause a loss of spin coherence in repeated measurements, known as \emph{dephasing effect}. The initial condition of the finite temperature of the state affects both the left and right arm of the trajectories as a common mode. Hence, the final contrast is temperature-independent. Sources of such noise may arise in fluctuations induced in the homogeneous bias magnetic field and gravitational gradient, for instance. We will discuss this below (sec.~\ref{sec:spin-indep}). 

On the other hand, spin-dependent fluctuations directly impact the individual arms by perturbing the trajectories in such a way that the imprint of the initial state of the temperature remains. Hence, such are detrimental towards the overall contrast and punished severely based on the noise spectrum. We will provide examples of such below (sec.~\ref{sec:spin-dep}). 
For both spin-independent and spin-dependent noises, we will assume Gaussian white and Lorentzian noise to illustrate how contrast loss is affected.

We will give a brief overview of the Stern-Gerlach setup in sec.~\ref{sec:SG-setup}, and then we will discuss the spin contrast by considering the finite-temperature effects in the initial state preparation; see sec.~\ref{sec:spin-contrast}. We will apply the techniques to both spin-independent and spin-dependent noise by taking two different power-spectral densities (PSD) of the noise. Finally, we will conclude our analysis by constraining the parameters for illustration~\footnote{Note that we will not consider the rotational effects of the nanoparticle in this paper. The spin-related Humpty-Dumpty problem has been extensively discussed in~\cite{Japha:2022phw,Zhou:2024pdl,Rizaldy:2024viw}. Indeed, we can also discuss this linear noise while including the rotational effects, but here, we will keep the study simple, and we wish to revisit this problem at later stages.}.

\section{Stern-Gerlach setup}\label{sec:SG-setup}

As a widely investigated Stern-Gerlach interferometer (SGI) model, we consider a diamagnetic nanoparticle that is levitated in a background magnetic field with a linear gradient.
For this system, the Hamiltonian can be written as~\cite{Wan16_GM,Scala13_GM,Pedernales:2020nmf,Marshman:2021wyk}:~\footnote{Note that we are ignoring gravity here, we are assuming that a strong diamagnetic trap can be created such that the motion is constrained only along a single direction orthogonal to Earth's gravity. For examples of such a traps, see~\cite{Hsu:2016,Elahi:2024dbb}.}
\begin{equation}
    {H}_0 = \frac{{\textbf{p}}^2}{2m} - \frac{\chi_\rho m }{2\mu_0}\textbf{B}^2+\mu \,{\textbf{S}}\cdot \textbf{B}\, ,
\end{equation}
where $\textbf{p}$ labels the momentum of the particle, $\chi_\rho$ is the mass magnetic susceptibility of the particle, $\mu_0$ is the vacuum magnetic permeability, $\mu$ is the magnetic moment of the spin.

The magnetic field consists of a homogeneous bias field $\vec{B}_0=B_0\hat{z}$ in the $z$ direction and a quadrupole field $-\eta z\hat{z} + \eta x\hat{x}$, both of which are orthogonal to Earth's gravity. Hence, the magnetic field is modelled as, see~\cite{Marshman:2019sne}
\begin{equation}\label{magneticfield}
\left\{ 
\begin{aligned}
    &B_z=B_0-\eta z\, ,\\
    &B_x= \eta x  \, ,
    \end{aligned}
\right.
\end{equation}
where $\eta$ is the constant magnetic gradient. 
The translational motion in the $x$ direction and in the $y$ direction (which is the direction of gravity) is trapped by the magneto-gravitational trap. 
By providing the right hierarchy in the trapping frequencies and assuming that the motion is strictly along the $z$ direction, we can build a one-dimensional SGI model~\footnote{We are assuming an ideal case where we take the initial condition of $x=0$. In reality, it will be extremely hard and this will require knowing the center-of-mass motion along $x,z$ directions extremely well. We will need to initiate the experiment at $x=0$, in which case there will be no displacement due to the external inhomogeneous magnetic field along this direction.}\cite{Elahi:2024dbb}, with the following Hamiltonian\cite{Scala13_GM} :
\begin{align}\label{SGIsetup}
    H_0 = &\frac{p_z^2}{2m} - \frac{\chi_\rho m \eta^2}{2\mu_0}z^2 + \left(\frac{\chi_\rho m B_0\eta}{\mu_0} - \mu\eta S_z  \right)z \nonumber\\
    &+\left(\frac{-\chi_\rho m B_0^2}{2\mu_0}+\mu S_zB_0\right) \, .
\end{align}
The nanoparticle with spin state $S_z=0$ can be levitated around the position $z=z_0$, namely
\begin{equation}
    z_0=\frac{B_0}{\eta}  \, .
\end{equation}
Therefore, the oscillation mode around $z_0$ can be quantized by
\begin{equation}
    {z} - z_0 = \sqrt{\frac{\hbar}{2m\omega}} ({a}+{a}^\dagger) \, ,\quad {p}_z=-i\sqrt{\frac{\hbar m\omega}{2}}({a}-{a}^\dagger) \, ,
\end{equation}
which leads to the Hamiltonian
\begin{equation}
    H_0=\hbar\omega {a}^\dagger {a} +\lambda S_z ({a}+{a}^\dagger) \, .
\end{equation}
The frequency $\omega$ and the coupling parameter $\lambda$ are given by
\begin{equation}
    \omega = \sqrt{\frac{-\chi_\rho}{\mu_0}}\,\eta \, ,\quad \lambda=\mu\eta \sqrt{\frac{\hbar}{2m\omega}} \, .
\end{equation}
The spatial superposition size of the SGI model is given by
\begin{equation}\label{superpositionsize}
    \delta z_{max}=\frac{4\lambda}{\hbar\omega} \Delta z \, ,\quad
    \Delta z\equiv \sqrt{\frac{\hbar}{2m\omega}} \, ,
\end{equation}
where $\Delta z$ represents the spatial width of the wave packet in the harmonic oscillator. Further, in this paper, we consider a linear noise term in the SGI setup, namely
\begin{equation}\label{hamiltonianSGI}
    H=\hbar\omega\, {a}^\dagger {a} +  \lambda S_z ({a}+{a}^\dagger) + H_{noise} \, .
\end{equation}
We will model the linear noise $H_{noise}$ by
\begin{align}\label{eq:Hnoise}
    H_{noise} = \Delta\lambda_1(t) ({a}+{a}^\dagger) + \Delta \lambda_2(t) S_z ({a}+{a}^\dagger) \, ,
\end{align}
where we name $\Delta\lambda_{(1,2)}(t)$ as spin-independent and spin-dependent noise, respectively.
The noise terms will arise from the small stochastic fluctuations of the experimental parameters and the interactions between the nanoparticle and the environment. For instance, considering the fluctuations of the bias magnetic field in the $z$ direction, we will get the spin-independent noise $\Delta\lambda_1(t)$. 
The spin-dependent noise term $\Delta\lambda_2(t)$ can arise from the magnetic field gradient's fluctuations or the spin axis direction wobbling because of the rigid body rotational dynamics of the particle, see~\cite{Perdriat:2024zje}. 

Besides, we assume that the spin state is stable in the SGI setup because of the following two reasons: (1) the spin flips caused by noise are neglected due to the existence of the magnetic field \cite{Marshman:2021wyk}, and (2) the decoherence of the spin state caused by the noise from experimental environments will not be considered since the time of single run of SGI should be shorter than the coherence time of the spin. Therefore, we only consider the noise model $H_{noise}$ that solely affects the motion of the particle's spatial degrees of freedom instead of the embedded spin. Next, we will consider the spin contrast loss caused by the noise model in our SGI setup.

We obtain the SGI setup \eqref{hamiltonianSGI} by considering a diamagnetic particle in a linear magnetic field background. Remarkably, the Hamiltonian \eqref{hamiltonianSGI} is generic for SGI models with any harmonically trapped particle. Therefore, our treatment and analysis are also applicable to other SGI protocols, such as a single atom/ion or a Bose-Einstein condensate of cold atoms in a magnetic or optical trap.


\section{Spin Contrast}\label{sec:spin-contrast}

\subsubsection{Humpty-Dumpty effect}

We will assume that the particle is initially trapped in a harmonic potential with frequency $\omega$, and that the spatial motion of the nanoparticle is cooled to a low temperature by cooling the center of mass motion, which depends on the trap, for instance, in diamagnetically levitated schemes; see~\cite{DUrso16_GM,Elahi:2024dbb}.

The initial spatial quantum state of the trapped nanoparticle is set to be in a coherent state $|\alpha\rangle$ of the trap, where $\alpha$ is a complex number. The particle wave packet $\bra{z}\alpha\rangle \propto exp[-(z\sqrt{m\omega/2\hbar}-\alpha)^2]$ is Gaussian shaped,
where $|z\rangle $ is the positional basis of the nanoparticle. Meanwhile, the initial spin state of the particle is prepared as a superposition of $S_z=+1$ and $S_z=-1$ state (denoted as $\lvert\uparrow\rangle$ and $\lvert\downarrow\rangle$ respectively), namely~\cite{Bose:2017nin}
\begin{equation}\label{unentangled}
    |s_0\rangle = \frac{1}{\sqrt{2}}(\lvert\uparrow\rangle+\lvert\downarrow\rangle)\, .
\end{equation}
Then, the full initial quantum state can be written as $\ket{\Psi_0}=\ket{\alpha}\otimes \ket{s_0}$.

In this case, considering the thermal distribution of $\alpha$ at finite temperature $T$, the density matrix operator of the spin-embedded particle can be described by~\cite{scala2013matter,Steiner:2024stq}
\begin{equation}\label{rho0}
    \rho_0 = \int \frac{d^2\alpha}{\pi}\frac{e^{-\frac{|\alpha|^2}{n}}}{n} \lvert\alpha\rangle\langle \alpha|\otimes \rho_{s0} \, ,
\end{equation}
where the thermal occupation number $n$ is defined by 
\begin{equation}
    n\equiv \frac{k_B T}{\hbar\omega} \, ,
\end{equation}
and, from Eq.~\eqref{unentangled}, the initial density matrix of spin is given by
\begin{equation}
    \rho_{s0}=\frac{1}{2} \begin{pmatrix} 1&1\\1&1 \end{pmatrix}\, .
\end{equation}
The time evolution of the density matrix is governed by the quantum Liouville equation 
\begin{equation}
    \rho(t)=U(t)\rho_0U^\dagger(t)\,,\,\,\,\,\, U(t)\equiv \exp\left(-\frac{i}{\hbar}\int_0^t dt' H(t') \right)\, .
\end{equation}
From the initial density matrix Eq.~(\ref{rho0}) and the Hamiltonian Eq.~(\ref{hamiltonianSGI}), we have
\begin{equation}\label{rhot}
  \rho(t)=\int \frac{d^2\alpha}{\pi}\frac{e^{-\frac{|\alpha|^2}{n}}}{n} \frac{1}{2} \begin{pmatrix}
  \lvert\psi_{L}\rangle\langle\psi_{L}|& \lvert\psi_{L}\rangle\langle\psi_{R}|\\
  \lvert\psi_{R}\rangle\langle\psi_{L}|&
  \lvert\psi_{R}\rangle\langle\psi_{R}|
  \end{pmatrix} ,
\end{equation}
note that this density matrix consists of the $2\cross2$ density matrix of the spin degrees of freedom and the infinite degrees of freedom from the evolution of the coherent state determined by the position and momentum; the latter is encoded in
the quantum states $\lvert\psi_{L, R}\rangle$ of the left and right arms of the SGI and are governed by the evolution equations
\begin{equation}\label{psiLpsiR}
\begin{aligned}
    \lvert\psi_L\rangle \equiv e^{-\frac{i}{\hbar}\int dt H_L}\lvert\alpha\rangle\ ,\quad
    \lvert\psi_R\rangle \equiv e^{-\frac{i}{\hbar}\int dt H_R}\lvert\alpha\rangle\ ,
\end{aligned}
\end{equation}
where the Hamiltonian $H_{L,R}$ represent the SGI Hamiltonian Eq.~\eqref{hamiltonianSGI} in the case of $S_z=\pm 1$, respectively. 
In the density matrix Eq.~\eqref{rhot}, the $\lvert\psi_{L, R}\rangle$ contains the information of their respective spatial trajectories, and the full quantum state of the spin-embedded particle is given by:  $\ket{\Psi}=(\ket{\psi_L}\otimes\ket{\uparrow}+\ket{\psi_R}\otimes\ket{\downarrow})/\sqrt{2}$.

Now, by taking the partial trace over the dynamics, i.e., degrees of freedom of the spatial trajectories, we, therefore, get (from Eq.~\eqref{rhot}) the traced density matrix of {\it just} the embedded spin:
\begin{equation}\label{rhos}
    \rho_s(t) \equiv \int dz \bra{z}\rho(t)\ket{z} =
     \frac{1}{2} \begin{pmatrix}
  1& \beta & \\
  \beta^* & 1 
\end{pmatrix}\, ,
\end{equation}
where $\ket{z}$ represents the complete basis of the spatial motional state of the particle. 

The complex number $\beta$ labels the overlap between the quantum state of the left and right arms of the interferometer, namely
\begin{align}
    \beta &\equiv \int \frac{d^2\alpha}{\pi}\frac{e^{-\frac{|\alpha|^2}{n}}}{n} \langle \psi_R\lvert \psi_L\rangle \nonumber\\
    & = \int\frac{d^2\alpha}{\pi}\frac{e^{-\frac{|\alpha|^2}{n}}}{n} \langle \alpha\lvert e^{\frac{i}{\hbar}\int dt H_R}e^{-\frac{i}{\hbar}\int dt H_L} \lvert\alpha\rangle \, .
\end{align}
The norm of the diagonal element $|\beta|$, usually denoted as $C$
$C= |\beta|$ is known as the \emph{spin contrast (or spin coherence)}; it represents the quantum coherence of the spin density matrix Eq.~(\ref{rhos}). 

The ideal case is that the two wave packets of the interferometric arms can match perfectly when we measure the spin, that is, $C=|\beta|=1$ so that the spin density matrix $\rho_s$ is pure and there is no decay of the off-diagonal elements (no coherence loss). 
However, if it is not possible to obtain $C=1$ due to some inevitable noise in a real experiment, then the contrast loss is inevitable. In the case of $|\langle \psi_R\lvert \psi_L\rangle|<1$, the spatial wave packet of the particle takes away the up/down information of the embedded spin, which means that the coherence of spin is partially or completely lost. This effect in SGI setups is precisely known as \emph{Humpty-Dumpty effect}, due to~\cite{Englert,Schwinger,Scully}.
The overlap of the parameter $\beta$ can generally be written as 
\begin{equation}\label{eq:beta}
    \beta \equiv C\, e^{i\delta\phi},
\end{equation}
where $\delta\phi$ is the \emph{interferometric phase} of the interferometer. 
Since the phase factor $e^{i\delta\phi}$ has norm one, it does not decohere the spin state in a single spin measurement. 
However, random run-to-run fluctuations of the relative phase $\delta\phi$ cause a loss of spin coherence in repeated measurements which becomes clear by averaging over the repeated runs, this is known as the \emph{dephasing effect}.

\subsubsection{Dephasing effects}

Considering repeated experiments, the statistical average of the spin density matrix Eq.~(\ref{rhos}) is given by:
\begin{equation}\label{arhos}
    \mathbb{E}[\rho_s]=
    \frac{1}{2} \begin{pmatrix}
  1& \mathbb{E}[\beta]  \\
  \mathbb{E}[\beta] & 1 
\end{pmatrix}\, ,
\end{equation}
where $\mathbb{E}[\cdot]$ denotes statistical average. Even in the case of $C=1$ (i.e. $\beta=e^{i\delta\phi}$), the statistical average of $e^{i\delta\phi}$ leads to an additional decay factor, given by, see~\cite{Milburn:2015}:
\begin{equation}
    \mathbb{E}[e^{i\delta\phi}]= e^{-\Gamma/2}e^{-i \mathbb{E}[\delta\phi]}\, , 
\end{equation}
where we assume that the value of $\delta\phi$ satisfies the Gaussian normal distribution and has the same statistical weight in each experimental run. The parameter $\Gamma$ is equal to the variance of $\delta\phi$, namely 
\begin{equation}\label{eq:generic-noise}
    \Gamma\equiv \mathbb{E}[(\delta\phi)^2]-(\mathbb{E}[\delta\phi])^2\, .
\end{equation}
Due to the dephasing effect, we are unable to completely extract the quantum information of the spin due to the randomness of the noise, which partially leads to spin contrast loss.

Considering both the Humpty-Dumpty and the dephasing effect, we can define an effective spin contrast $\widetilde{C}$, which contains both the information about the \textit{Humpty-Dumpty} (in $C$) and the \textit{dephasing} (in $\mathbb{E}[e^{i\delta\phi}]$), given by the norm of the $\mathbb{E}[\beta]$ for repeated SGI measurements, namely:
\begin{equation}\label{contrast}
    \widetilde{C}\equiv \abs{\mathbb{E}[\beta]} .
\end{equation}

In addition, there will be another source of contrast loss in SGI experiments as a result of the effects of decoherence. The environment will take away the quantum information of the interferometric system due to interactions between them, such as air molecule scattering and black body radiation. These will contribute to a damping factor $e^{-\gamma_d\,t}$ on spin coherence, where $\gamma_d$ is the \emph{decoherence rate}, see~\cite{bassireview,Schlosshauer:2014pgr,Hornberger_2012,ORI11_GM}. The effect has been investigated in detail in many previous works~\cite{bassi2013models,Schut:2024lgp,Rijavec:2020qxd,Schut:2023eux,Schut:2021svd}. As mentioned in the section. II, the time scale of SGI is set within the spin coherence time, so we will not consider the decoherence effect here.

In the following sections, we will investigate the impact of spin-independent and spin-dependent noisy Hamiltonian $H_{noise}$, separately, on the effective spin contrast in the case of finite temperature.


\section{spin-independent noise}\label{sec:spin-indep}

We first consider the spin-independent noise that the noise term in Hamiltonian takes the form $${H}_{noise}(t)=\Delta\lambda(t) ({a}+{a}^\dagger),$$ where $\Delta\lambda(t)$ represents a time-dependent noise acting during interferometry. Therefore, the nanoparticle is now governed by the Hamiltonian, including the noise:
\begin{equation}
    {H}=\hbar\omega {a}^\dagger {a} + (\lambda S_z + \Delta\lambda(t)) ({a}+{a}^\dagger)\, .
\end{equation}

\subsection{Humpty-Dumpty effect}\label{subsec:HD}

The time evolution Eq.~\eqref{psiLpsiR} of the wave functions of the left and right arms of the trajectories can be solved by the quantum forced harmonic oscillator model, see appendix \ref{app:a}, where the evolution operator can be parameterized by a phase $\varphi_\pm$  and a displacement parameter $\zeta_\pm$, namely, 
\begin{eqnarray}\label{psiLpsiRSI}
     \lvert\psi_L\rangle= e^{i\int dt H_L}\ket{\alpha} \equiv U_0 e^{i\varphi_+} D(\zeta_+)|\alpha\rangle \,,\nonumber \\
     \lvert\psi_R\rangle= e^{i\int dt H_R}\ket{\alpha} \equiv U_0 e^{i\varphi_-} D(\zeta_-)|\alpha\rangle \, ,
\end{eqnarray}
where $U_0\equiv e^{-i\omega t {a}^\dagger {a}}$ is a unitary operator and $D(\zeta)\equiv \exp (\zeta {a}^\dagger -\zeta^* {a})$ is the (time-dependent) displacement operator. When calculating the overlap, the unitary operator will be cancelled: $\langle \psi_L|\psi_R\rangle$. 

There will be fluctuations arising from the phase $\varphi(t)$, and from the displacement parameter $\zeta$. The former will give rise to random fluctuations in phase, i.e. dephasing, which can computed by Eq.~(\ref{contrast}). The latter fluctuations will amount to the well-known Humpty-Dumpty problem~\cite{Englert,Scully,Schwinger}.

The solution of the displacement parameter $\zeta_\pm$ (see Eq.~\eqref{eq:zeta}) is given by:
\begin{equation}
    \zeta_\pm (t) = \mp u (e^{i\omega t}-1) -i\omega\int_0^t dt' \Delta u(t')e^{i\omega(t'-t)}\, ,
\end{equation}
where we define the dimensionless parameter, $u$, and the dimensionless noise $\Delta u(t)$ as:
\begin{equation}
    u\equiv \frac{\lambda}{\hbar\omega}\, ,\quad\Delta u(t)\equiv\frac{\Delta\lambda(t)}{\hbar\omega} \, .
\end{equation}
Taking the evolution time as exactly the period of the harmonic trap, namely, $\omega t=2\pi$, we have 
\begin{equation}\label{zetaSI}
    \zeta_+\left(\frac{2\pi}{\omega} \right)=\zeta_-\left(\frac{2\pi}{\omega} \right)=-i\omega\int_0^{{2\pi}/{\omega}} dt' \Delta u(t')e^{i\omega t'} \, .
\end{equation}
Equation~\eqref{zetaSI} indicates that there is no mismatch in the location and momentum of the wave packets for an initial thermal state $\lvert\alpha\rangle$ since $ \zeta_+\left({2\pi}/{\omega} \right)=\zeta_-\left({2\pi}/{\omega} \right)$, i.e., there is no relative displacement between the wave packets in the left and right arm of the interferometer. In the context of Eq.~\eqref{eq:generic-noise}, we therefore find $\abs{C} = 1$.
Hence, the linear spin-independent noise does not cause the Humpty-Dumpty effect on the spin coherence loss. Note that there is no contrast loss due to the initial motional state of the nanoparticle in a harmonic trap either. This corroborates earlier results, although shown in a different setting; see~\cite{Scala13_GM}.

\subsection{Dephasing due to spin-independent noise}

However, despite having no Humpty-Dumpty problem in this case, we will still incur dephasing due to the spin-independent noise. We will now elaborate on this aspect. Note that in Eq.~\eqref{psiLpsiRSI} there is also the path-dependent phase $\varphi_\pm$, which is derived in Appendix~\ref{app:a} to be: 
\begin{align}\label{varphipm}
    \varphi_\pm = \omega^2 \int_0^{{2\pi}/{\omega}}dt\int_0^{t} dt'\, & [\pm u +\Delta u(t)][\pm u +\Delta u(t')] \nonumber\\
    & \times {\rm sin}(\omega(t-t')) \, ,
\end{align}
the phase difference between the two arms of the matter-wave interferometer,  $\delta\varphi(t)= \varphi_+(t) - \varphi_-(t)$, is given by
\begin{equation}\label{deltaphiSI}
    \delta\varphi=2u\omega^2 \int_0^{\frac{2\pi}{\omega}}dt\int_0^{t} dt' \, [\Delta u(t)+\Delta u(t')]{\rm sin}(\omega(t-t'))\, ,
\end{equation}
The spin-independent noise leads to time-dependent random fluctuations in the phase difference between the two arms, hence leading to dephasing. 
From Eqs.~\eqref{psiLpsiRSI},~\eqref{zetaSI},~\eqref{deltaphiSI}, we have $\langle\psi_L|\psi_R\rangle=e^{i\delta\varphi}$. Hence, the overlap parameter $\beta$ (Eq.~\eqref{eq:beta}) reads
\begin{equation}
    \beta = \int \frac{d^2\alpha}{\pi}\frac{e^{-\frac{|\alpha|^2}{n}}}{n} e^{i\delta\varphi} = e^{i\delta\varphi} \, .
\end{equation}
Therefore, the effective spin contrast is reduced by the dephasing effect, described by $\mathbb{E}[\beta]=e^{-\mathbb{E}[\delta\varphi^2]/2}$. Note that here, we considered the noise $\Delta u$ with zero average value, namely $$\mathbb{E}[\Delta u]=\mathbb{E}[\delta\varphi]=0.$$
This can often be achieved experimentally by performing a control experiment.

\begin{figure}[ht]
    \centering
    \includegraphics[width=0.9\linewidth]{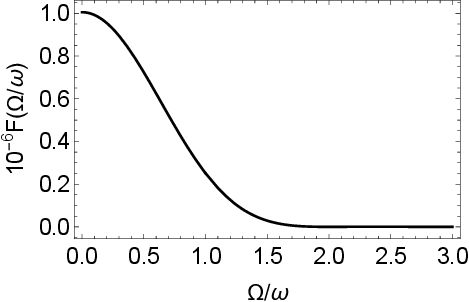}
    \caption{The transfer function for the variance of the phase difference due to linear spin-independent noise; see Eq.~\eqref{transferfunc1}. We set $u=100$ in this plot. See Appendix~\ref{app:b} for the derivation of the transfer function.}
    \label{transferfunction1}
\end{figure}

Let us determine the spin contrast loss due to the dephasing effect for general noise models. Applying the solution \eqref{deltaphiSI}, the variance of $\delta\varphi$ is
\begin{align}\label{phasevariance}
    \mathbb{E}[\delta\varphi^2]= & 4u^2\omega^4 \int_0^{\frac{2\pi}{\omega}} dt_1 \int_0^{\frac{2\pi}{\omega}} dt_2 \int_0^{t_1} dt_1' \int_0^{t_2} dt_2' \nonumber\\
    &\mathbb{E}[[\Delta u(t_1)+\Delta u(t_1')][\Delta u(t_2)+\Delta u(t_2')]] \nonumber\\
    &\times{\rm sin}\omega(t_1-t_1'){\rm sin}\omega(t_2-t_2')\, .
\end{align}
According to Wiener-Khinchin theorem~\cite{Wiener:1930,Khintchine1934}, the spectral decomposition of the autocorrelation function of a stationary random process is given by the \emph{power spectral density (PSD)} of the noise\footnote{Here we consider a stationary noise such that the PSD function does not change during the total experimental time, comprising all the repeated SGI runs. 
A cutoff $\Omega_\text{min}$ of the noise frequency $\Omega$ usually is taken in the integral of Eq.~\eqref{PSD} since the total experiment time is finite and the noise is sampled as a discrete process, which leads to a finite frequency resolution. However, we take $\Omega_\text{min}\rightarrow 0$ in this work because we consider many experimental runs so that we have $\Omega_\text{min}\ll \omega$. Moreover, taking the small cutoff $\Omega_\text{min}$ to zero does not affect the calculation of spin contrast too much because the transfer functions are finite, see Fig.~\ref{transferfunction1} and Fig.~\ref{transferfunction2}.}, i.e.
\begin{equation}\label{PSD}
    \mathbb{E}[\Delta u (t)\Delta u(t')]=\int_0^\infty \frac{d\Omega}{2\pi} \ S_{\Delta u}(\Omega)\, e^{i\Omega (t-t')} \, ,
\end{equation}
where $\Omega$ is the frequency of the noise $\Delta u$, and $S_{\Delta u}(\Omega)$ denotes the PSD function of the noise. Therefore, from the Eqs.~\eqref{phasevariance},~\eqref{PSD}, we get the variance of $\delta\varphi$ (see the derivation in Appendix~\ref{app:b}):
\begin{equation}\label{dephase}
    \Gamma=\mathbb{E}[\delta\varphi^2]=\int_0^\infty S_{\Delta u}(\Omega) F(\Omega/\omega)\, d\Omega \, , 
\end{equation}
where $F(\Omega/\omega)$ is known as the \emph{transfer function}~\cite{Toros:2020dbf,Wu:2022rdv,Schut:2023tce}, which is given by 
\begin{equation}\label{transferfunc1}
    F(x)= \frac{32u^2}{\pi} \frac{{\rm sin}^2\left(\pi  x \right)}{(x^3-x)^2}\,,~~~~x=\frac{\Omega}{\omega}\,.
\end{equation}
From Fig.~\ref{transferfunction1}, we can see that the dephasing effect is mainly affected by low-frequency noise in the regime $\Omega\lesssim \omega$.

We will now consider two examples of noise purely for illustration. One white and the other Lorentzian noise will be used to analyze the dephasing. The latter is very common in the sense that such noise may appear due to fluctuations in the current, which might give rise to fluctuations in the bias magnetic field.

\begin{itemize}

\item{White Noise}:
As a first example, we will take the continuous-time domain Gaussian white noise model~\cite{Milburn:2015}
\begin{equation}
S_{\Delta u}^W= \frac{\sigma^2}{\omega}\,,
\end{equation}
where $\sigma^2$ is a constant and represents the variance of noise, $\Delta u$, measured in the interferometric time, scale $t=2\pi/\omega$. Therefore, by substituting in Eq.~(\ref{dephase}) with the help of Eq.~(\ref{transferfunc1}), we can obtain the dephasing parameter
\begin{equation}
    \quad \Gamma_W=\frac{32u^2}{\pi}\int_0^\infty \frac{\sigma^2}{\omega} F\left(\frac{\Omega}{\omega} \right) \, d\Omega=24\pi u^2\sigma^2 \, ,
\end{equation}
Then, by using Eq.~(\ref{contrast}), we  obtain  the effective spin contrast, given by:
\begin{equation}
    \widetilde{C}=\mathbb{E}[e^{i\delta\varphi}]=e^{-\Gamma_W/2}=e^{-12\pi u^2\sigma^2}\, .
\end{equation}
Here, we have used the result of sec.~\ref{subsec:HD} that $\abs{C}=1$ such that only the dephasing plays a role in determining the effective spin contrast.
Note that the spin contrast is decaying exponentially, and it is sensitive to the dimensionless coupling
$$u\equiv\frac{\lambda}{\hbar\omega}=\mu\sqrt{\frac{- \mu_0}{2 \hbar\omega m\chi_\rho}},$$ 
The Gaussian white noise also depends on the constant $\sigma^2$ which represents the variance of $\Delta u$; we can see in Fig.~\ref{contrast1} (the black dashed line) how the effective spin contrast evolves concerning $\sigma$, which determines the magnitude of the Gaussian white noise. To obtain a large spin contrast, the deviation $\sigma$ of the fluctuation $\Delta u(t)$ should be minimized to the limit $\sigma \ll 1/u = \hbar\omega/\lambda$. Recalling Eq.~(\ref{superpositionsize}), $\delta z_{max}=4u\Delta z$. We see that to minimize contrast loss; we will need: $\sigma \ll 4\Delta z/\delta z_{max}$. Hence, large superposition size is less favourable due to contrast loss induced by dephasing.


\begin{figure}[ht]
    \centering
    \includegraphics[width=\linewidth]{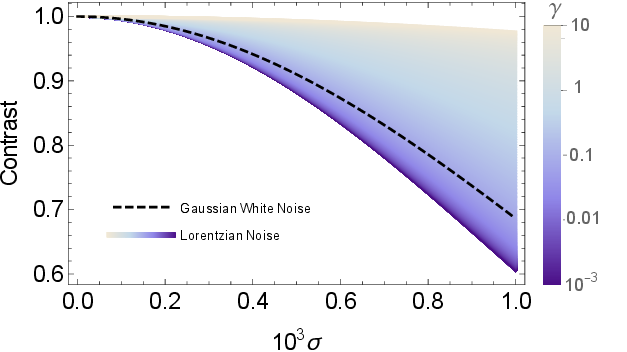}
    \caption{Spin contrast loss due to the spin-independent Gaussian white noise (dashed curve) and the Lorentzian noise (shaded region). Here, we set $u=100$. We can consider a single NV-center embedded nanodiamond with $\chi_\rho\sim -6.2\times 10^{-9}{\rm m^3/kg}$, $m\sim 10^{-17}$~kg, $\omega\sim 10^3$~Hz, which gives $u\sim 10^2$.  Note that the spin contrast is large for $\gamma \sim 10$. For large $\gamma$, the Lorentzian PSD is nearly constant $\propto \gamma^{-1}$ in a low-frequency regime while the transfer function maximises for smaller $\Omega$, giving rise to a large spin contrast.} 
    \label{contrast1}
\end{figure}


\item{Lorentzian noise}:
For the purpose of illustration, we can also consider $\Delta u(t)$ to have a Lorentzian power spectrum. The PSD of the Lorentzian noise is modelled by\cite{Horn:1981}
\begin{equation}\label{Lorentz-N}
    S_{\Delta u}^L (\Omega) \equiv \frac{\sigma^2}{\omega} \frac{2\gamma/\pi}{[(\Omega-\Omega_0)/\omega]^2+\gamma^2} \, ,
\end{equation}
where $\Omega_0$ is the noise resonance frequency, the parameter $\gamma$ indicates the linewidth (in unit of $\omega$) of the PSD distribution and $\sigma^2$ is the variance of the Lorentzian noise $\Delta u$.~\footnote{ Note that this notation $\gamma$ should not be confused by the decoherence rate. In this paper, we denote the decoherence rate by $\gamma_d$; see the discussion after Eq.~(\ref{contrast}).}. In the following discussion, we set $\Omega_0=0$ because that case is common in solids and circuits.
\begin{figure}[bpt]
    \centering
    \includegraphics[width=\linewidth]{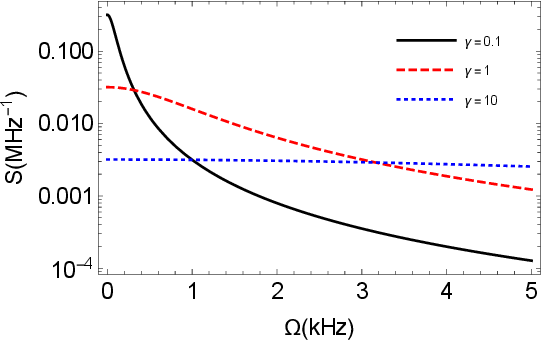}
    \caption{PSD of the Lorentzian noise model given by  Eq.~\eqref{Lorentz-N}. In this plot, we set the power of the noise $\sigma=10^{-2}$, $\Omega_0=0$, and $\omega=1$~kHz. We can see that taking the smaller value of the parameter $\gamma$, the amplitude of the PSD is more dominated by the low-frequency regime. }
    \label{pdfLorentzian}
\end{figure}

For this Lorentzian noise PSD, we can compute the dephasing parameter like before, and we obtain:
\begin{equation}
    \quad\quad\Gamma_L=\int_0^\infty \frac{\sigma^2}{\omega} \frac{2\gamma/\pi}{(\Omega/\omega)^2+\gamma^2} F\left(\frac{\Omega}{\omega} \right) \, d\Omega\,. 
\end{equation}
We will compute the spin contrast numerically, shown below in Fig.~\ref{contrast1}.

The spin contrast decay as a function of $\sigma$ is plotted in Fig.~\ref{contrast1} as the shaded region (as a function of $\gamma$). Besides, we see that a large $\gamma$ favours spin contrast because, for the PSD function \eqref{Lorentz-N}, the power of noise contributes less in the low-frequency regime when $\gamma$ is large, see Fig.~\ref{pdfLorentzian}.
\end{itemize}

In our case, the dephasing is not affected by the finite temperature of the initial state. The fluctuations are solely in the phases of the two paths $\varphi_{\pm}$ and not in the displacement parameter $\zeta_{\pm}$. This is because in the harmonic trap, the evolution of the global phase $\varphi_{\pm}$, see equation Eq.~(\ref{varphipm}), is independent of the initial thermal distribution of $\alpha$.


\section{spin-dependent noise}\label{sec:spin-dep}

Let us now  consider a spin-dependent noise term $H_{noise}=\Delta\lambda(t)S_z({a}+{a}^\dagger)$ in the Hamiltonian, namely
\begin{equation}
    H(t) = \hbar\omega {a}^\dagger {a} + (\lambda + \Delta\lambda(t))S_z ({a}+{a}^\dagger)\, .
\end{equation}
The left and right wave functions, i.e. the two arms of the SGI, are also coherent states described by the form Eq.~(\ref{psiLpsiRSI}).  We can solve the above Hamiltonian exactly by following the interaction picture in quantum mechanics; see Appendix~\ref{app:a}. 

According to Eqs.~(\ref{zetatheta},~\ref{evolutionoperator}) in appendix~\ref{app:a},  and taking $t=2\pi/\omega$, the parameter $\zeta_{\pm}$ and $\varphi_\pm$ can be solved by:
\begin{align}
    \zeta_{\pm}&=\mp i\omega \int_0^{\frac{2\pi}{\omega}} dt \Delta u(t)e^{i\omega t}\, , && \label{zetapm} \\
    \varphi_+&=\varphi_-=\omega^2\int_0^{\frac{2\pi}{\omega}} dt\int_0^{t}dt' &&\hspace{-3mm}[u+\Delta u(t)][u+\Delta u(t')] \nonumber\\&& & \times {\rm sin}(\omega(t-t')) \, .
\end{align}
Remarkably, contrary to the spin-independent noise, we can see that the linear spin-dependent noise only causes the Humpty-Dumpty effect instead of the dephasing because 
$\varphi_{-}=\varphi_{+}$, there are no phase fluctuations in the left and right arms of the SGI in this case.
This is because, in this case, the term $u + \Delta u$ has an overall spin-dependent sign, which is cancelled, while in the spin-independent case, $u -\Delta u$ had a relative spin-dependent sign, which could not be cancelled.
However, the displacement parameter $\zeta_{-}\neq \zeta_{+}$, and hence there will be the Humpty-Dumpty problem.

Therefore, based on Eqs.~ (\ref{psiLpsiRSI},~\ref{zetapm}), the overlap between the left and right wave function is given by:
\begin{align}
    \langle\psi_R\lvert\psi_L\rangle &=\langle\alpha|D^\dagger(\zeta_-)D(\zeta_+)|\alpha\rangle \nonumber\\
    &= e^{-\frac{1}{2}|\delta\zeta|^2} e^{\delta\zeta\cdot \alpha^*-\delta\zeta^*\alpha } \, ,
\end{align}
where we have used the expression of the displacement operator given in Eq.~\eqref{Ui} (see appendix~\ref{app:b} for more details).
The mismatch $\delta\zeta$ is given by:
\begin{equation}
    \delta\zeta\equiv\zeta_+-\zeta_-=-2i\int_0^{{2\pi}/{\omega}} dt \Delta u(t)e^{i\omega t}\, .
\end{equation}
Therefore, after taking the thermal average for the initial coherent state $\lvert\alpha\rangle$, we have
\begin{equation}\label{beta}
    \beta = \int \frac{d^2\alpha}{\pi}\frac{e^{-\frac{|\alpha|^2}{n}}}{n} \langle \psi_R\lvert \psi_L\rangle  = e^{-(\frac{1}{2}+n)|\delta\zeta|^2}\, .
\end{equation}
Then, the statistical average of $\beta$ is given by:
\begin{equation}
    \mathbb{E}[\beta]=\mathbb{E}\left[ e^{-(\frac{1}{2}+n)[\Re^2(\delta\zeta)+\Im^2(\delta\zeta)]} \right] \, ,
\end{equation}
where
\begin{align}
    &\Re(\delta\zeta)=2\omega \int_0^{\frac{2\pi}{\omega}} dt \Delta u(t) \sin(\omega t) \, ,\nonumber\\
    &\Im(\delta\zeta)= 2\omega \int_0^{\frac{2\pi}{\omega}} dt \Delta u(t) \cos(\omega t) \, ,
\end{align}
represent the location and the momentum mismatch between the left and the right trajectory of the SGI, respectively.
Using the equation~\footnote{Note that $\mathbb{E}[e^{-a\cdot x^2}]= \int_{-\infty}^{\infty} e^{-ax^2}\frac{1}{\sqrt{2\pi \mathbb{E}[x^2]}}e^{-x^2/2\mathbb{E}[x^2]} dx$, which gives $\mathbb{E}[e^{-a\cdot x^2}]= \sqrt{\frac{1}{1+2a\mathbb{E}[x^2]}}$.
}:
$$\mathbb{E}[e^{-a\cdot x^2}]=(1+2a\mathbb{E}[x^2])^{-\frac{1}{2}}\,,$$ 
which holds when $x$ satisfies a normal distribution with zero mean, we obtain the statistical average of $\beta$, as:
\begin{align}\label{Ebeta}
&\mathbb{E}\left[ e^{-(\frac{1}{2}+n)[\Re^2(\delta\zeta)+\Im^2(\delta\zeta)]} \right] \nonumber\\
&=\left([1+(1+2n)\mathbb{E}[\Re^2(\delta\zeta)]] [1+(1+2n)\mathbb{E}[\Im^2(\delta\zeta)]]\right)^{-\frac{1}{2}} \, .
\end{align}
The variance of the real part of $\delta\zeta$ is given by
\begin{align}\label{ERedeltaalpha}
    &\mathbb{E}[\Re^2(\delta\zeta)] \nonumber\\
    &=4\omega^2\mathbb{E}\left[\int_0^{\frac{2\pi}{\omega}}dt\int_0^{\frac{2\pi}{\omega}}dt' \Delta u(t)\Delta u(t'){\rm sin}(\omega t){\rm \sin}(\omega t')\right] \nonumber\\
    &=4\omega^2\int_0^{\frac{2\pi}{\omega}}dt\int_0^{\frac{2\pi}{\omega}}dt' \mathbb{E}[\Delta u(t)\Delta u(t')]{\rm sin}(\omega t){\rm sin}(\omega t')\nonumber\\
    &= \int_0^\infty d\Omega\, S_{\Delta u}(\Omega)F_{re}(\Omega)\, ,
\end{align}
where the transfer function of $\mathbb{E}[\Re^2(\delta\zeta)]$ is:
\begin{align}\label{Fre}
    F_{re}(\Omega)&= 4\omega^2\int_0^{\frac{2\pi}{\omega}}dt\int_0^{\frac{2\pi}{\omega}}dt' e^{i\Omega(t-t')} {\rm sin}(\omega t){\rm sin}(\omega t') \nonumber\\
    & = \frac{16\, {\rm sin}^2(\pi x)}{(x^2-1)^2}\, ,\quad {\rm where}\ x=\Omega/\omega \, .
\end{align}
Similarly, we can obtain the variance of the imaginary part of $\delta\zeta$:
\begin{equation}\label{EImdeltaalpha}
    \mathbb{E}[\Im^2(\delta\zeta)]=\int_0^\infty d\Omega\, S_{\Delta u}(\Omega)F_{im}(\Omega)\, ,
\end{equation}
where the transfer function of $\mathbb{E}[\Im^2(\delta\zeta)]$ is::
\begin{align}\label{Fim}
    F_{im}(\Omega)&= 4\omega^2\int_0^{\frac{2\pi}{\omega}}dt\int_0^{\frac{2\pi}{\omega}}dt' e^{i\Omega(t-t')} {\rm cos}(\omega t){\rm cos}(\omega t') \nonumber\\
    & = \frac{16\, {\rm sin}^2(\pi x)}{(x-\frac{1}{x})^2}\, ,\quad {\rm where}\ x=\Omega/\omega \, .
\end{align}

\begin{figure}
    \centering
    \includegraphics[width=0.8\linewidth]{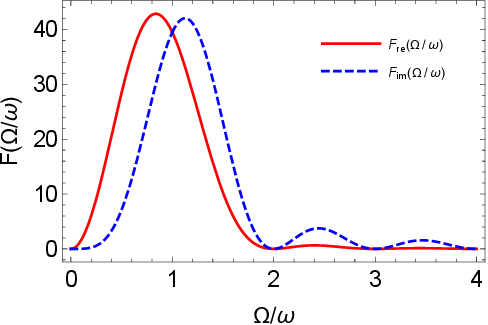}
    \caption{The transfer functions, denoted as $F_{re}(\Omega)$ and $F_{im}(\Omega)$, for the variance of the wave packet's location and momentum mismatch due to the spin-dependent noise.}
    \label{transferfunction2}
\end{figure}

The transfer functions of the variance of $\delta\zeta$ are shown in Fig.~\ref{transferfunction2}. One can see that the spin contrast loss induced by spin-dependent noises is mainly dependent on the noise with the frequency $\Omega\approx \omega$.

\begin{itemize}

\item{White noise}: By considering Gaussian white noise, namely~\cite{Milburn:2015} 
$$S_{\Delta u}=\sigma^2/\omega, $$ 
we can obtain the variance of the $\delta\zeta$ from Eqs.~(\ref{ERedeltaalpha},\ref{Fre}) and (\ref{EImdeltaalpha},\ref{Fim})
\begin{equation}\label{Edeltaalpha2}
    \mathbb{E}[\Re^2(\delta\zeta)]=\mathbb{E}[\Im^2(\delta\zeta)]=4\pi^2\sigma^2 \, .
\end{equation}
Therefore, employing (\ref{Edeltaalpha2}) to equation (\ref{Ebeta}),  we obtain the effective spin contrast
\begin{equation}
    \tilde{C}=\mathbb{E}[\beta]= \frac{1}{1+(4+8n)\pi^2\sigma^2} \, .
\end{equation}
We see that as $n \gg 1$, the contrast loss will be significant as compared to the case of zero temperature case, i.e. $n=0$. Figs.~\ref{contrast2} depict our scenarios for a specific value of $n=100$ in Fig.~\ref{contrast2}(a) w.r.t $\sigma$, and for a fixed value of $\sigma=10^{-2}$  in Fig.~\ref{contrast2}(b) w.r.t. varying $n$.

\item{Lorentzian noise}: 
We will also consider other PSD distributions of noise $\Delta u$, such as given by the Lorentzian noise in Eq.~(\ref{Lorentz-N}), which we then solve numerically for 
\begin{align}
    \qq{}&\mathbb{E}[\Re^2(\delta\zeta)]=\int_0^\infty \frac{\sigma^2}{\omega} \frac{2\gamma/\pi}{(\Omega/\omega)^2+\gamma^2} F_{re}\left(\frac{\Omega}{\omega} \right) \, d\Omega\nonumber\\
    \qq{}&\mathbb{E}[\Im^2(\delta\zeta)]=\int_0^\infty \frac{\sigma^2}{\omega} \frac{2\gamma/\pi}{(\Omega/\omega)^2+\gamma^2} F_{im}\left(\frac{\Omega}{\omega} \right) \, d\Omega \, , \nonumber
\end{align}
to find the effective spin contrast via Eq.~\eqref{Ebeta}. 
Fig.~\ref{contrast2}(c) and Fig.~\ref{contrast2}(d) show how the spin contrast decays with increasing noise variance and temperature, respectively. The results are depicted in Fig.~\ref{contrast2}(c) and Fig.~\ref{contrast2}(d) for varying $\sigma$, occupation number $n$ and $\gamma$. We can see that by increasing $\sigma$ and $n$, the Humpty-Dumpty problem becomes worse and the contrast decays. Remarkably, compared to the $\gamma\approx1$ case, the spin contrast is insensitive to Lorentzian noise with a too-small and large value of $\gamma$. In addition, for the more general Lorentzian noise spectrum in Eq.~\eqref{Lorentz-N} with nonzero $\Omega_0$, notice that the spin contrast will be lost mostly when $\Omega_0$ is close to $\omega$ (the trap frequency).


\begin{figure}[ht]
    \centering
    \includegraphics[width=\linewidth]{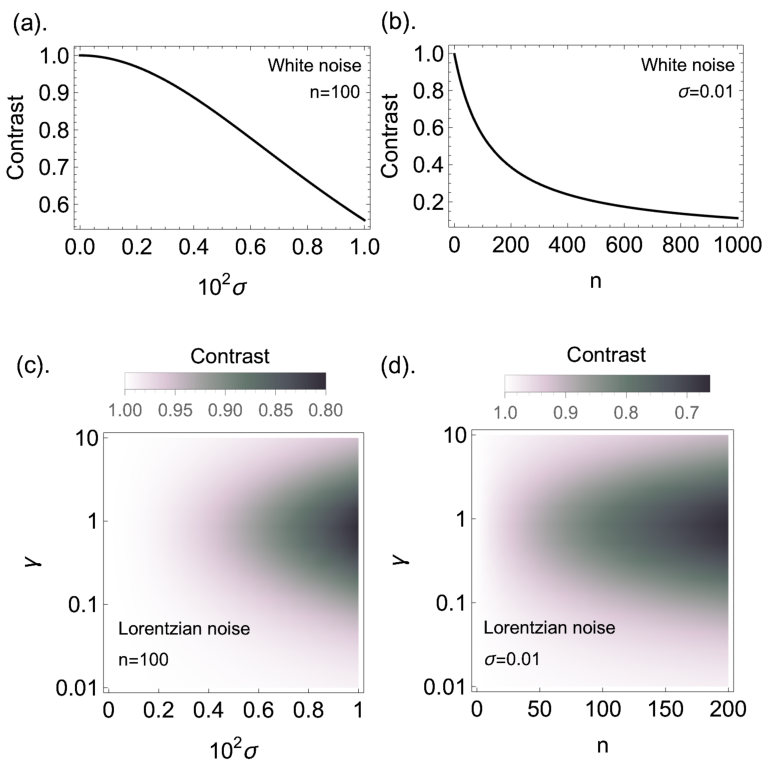}
    \caption{Plots (a) and (b) show the spin contrast loss as a function of the fluctuation amplitude $\sigma$ and the temperature $n$ for spin-dependent Gaussian white noise. Plots (c) and (d) show the spin contrast as a function of $\sigma$ and $n$ for spin-dependent Lorentzian noise. Here we set the temperature as $n=100$ ($T\approx 5\,\mu$K for a harmonic trap with frequency $\omega=1$~kHz) in subfigure (a) and (c), and set $\sigma=10^{-2}$ in subfigure (b) and (d). Note that the contrast is independent of the parameters of the SGI models, such as the magnetic field, its gradient, and the frequency of the trap. Furthermore, note that the contrast decreases more significantly when $\gamma\approx1$ in the case of Lorentzian noise. This is because the PSD (Fig.~\ref{pdfLorentzian}) of the Lorentzian noise is dominated by the low-frequency regime when $\gamma$ is small and contributes more power in the high-frequency regime when $\gamma$ is large, while the transfer function given by Fig.~\ref{transferfunction2} vanishes in the low- and high-frequency regimes.}
    \label{contrast2}
\end{figure}



\end{itemize}


\section{Conclusions}

In this work, we investigated the spin coherence loss resulting from random time-dependent phase fluctuations and the mismatch in the classical and quantum overlap of the wavepackets of the left and right arm of the matter wave interferometer. The former is known as dephasing, and the latter is known as the famous "Humpty-Dumpty" problem due to Englert, Scully, and Schwinger~\cite{Englert,Scully,Schwinger}. We investigated these two types of fluctuations for spin-independent and spin-dependent sources of linear noise terms in a harmonic oscillator potential. Here, we assumed that the spin is embedded in a trapped particle, and it is responsible for creating a macroscopic quantum superposition in an inhomogeneous magnetic field of the Stern-Gerlach interferometer, given by Eq.~\eqref{magneticfield}. The initial state of the harmonic oscillator is in a thermal ensemble. For illustration, we considered both white and Lorentzian PSDs to model the linear spin-independent/dependent noise terms.  

In this paper, we found that the linear spin-\textit{independent} noise causes a dephasing effect, in which the spin contrast exponentially decays with the growth of the noise variance and the particle's superposition size for the white/Lorentzian noise model. Moreover, the spin contrast loss caused by dephasing does not get worse due to the initial thermal motion in the harmonic trap. The dephasing effect tends to vanish in the large $\gamma$ limit for the Lorentzian noise model because the dephasing effect is solely sensitive to the low-frequency regime of noise; however, this is model-dependent from the perspective of creating the macroscopic quantum superposition and the origin of the white/Lorentzian noise. The procedure presented here can be applied to specific trajectory models. For this spin-independent noise, there is no Humpty-Dumpty effect, meaning that there is no loss in spin contrast due to the mismatch in the trajectories and the quantum overlap of the respective wavepackets. 

For the linear spin-\textit{dependent} noise, we found that this kind of noise results in the Humpty-Dumpty problem, leading to the mismatch of the location and the momentum of the left and right arms of the wavepackets, while there is no dephasing effect that causes loss of coherence. The Humpty-Dumpty effect is sensitive to noise with frequencies near the harmonic trap frequency. In this case, the spin contrast is inversely proportional to the variance of the noise and the temperature of the thermal motion. Moreover, the Humpty-Dumpty effect due to linear noise is independent of the parameters of the SGI models such as $\omega$, the magnetic field, and its gradient. However, the PSD of the noise, such as the parameter $\gamma$ of Lorentzian noise, can be model-dependent from the perspective of creating the superposition and the source of noise.

\begin{table}[t]
\centering
\caption{ As an example, we show tolerable standard deviation of spin-independent/dependent noise in an SGI setup for a superposition of size $\delta z\sim 50$~nm for a spin contrast $\tilde{C}>0.95$.}
\label{table1}
\renewcommand{\arraystretch}{1.2}
\setlength\tabcolsep{10pt}
\begin{tabular}{cc}
   \toprule
   Parameters & Values \\
   \midrule
   Mass of nanodiamond, $m$ & $10^{-17}$~kg \\
   Magnetic gradient, $\eta$  &  $1.4\times10^4$~T/m  \\
   Trap frequency, $\omega$  & $1$~kHz \\
   Superposition size, $\delta z$ & $\sim50$~nm \\
   Initial wave packet width, $\Delta z$ & $\sim7\times10^{-2}$~nm \\ 
   Initial temperature, $T$ &  $\sim5\,\mu$K ($n=100$)\\
   \makecell{ Tolerable standard deviation of  \\ spin-independent noise, $\sigma$} & $\lesssim2\times 10^{-4}$ \\
   \makecell{ Tolerable standard deviation of \\ spin-dependent noise, $\sigma$}&  $\lesssim2.5\times 10^{-3}$\\
   \bottomrule
\end{tabular}
\end{table}

Apart from the above generic properties of the impacts of noise on SGI, our work indicates how much the initial temperature of the particle's motion and the deviation of noise can be tolerated to maintain enough spin contrast. As an example, we consider a nanodiamond with mass $10^{-17}$kg in a $1$kHz magnetic trap as a one-dimensional interferometer. The tolerable amplitude $\sigma$ of the spin-independent/dependent noise $\Delta\lambda/(\hbar\omega)$ is shown in Table \ref{table1}. Our analysis can be the reference for the cooling of the initial state and noise control in future SGI experiments.

It will be worthwhile to do a similar noise analysis for more elaborate superposition creation models to see how modifying the initial condition affects the spin contrast for the matter-wave interferometers in Stern-Gerlach setups. For SGI with a massive particle, it will also be pertinent to perform the analysis including the rotation of the particle. As performed in Refs.~\cite{Zhou:2024pdl,Rizaldy:2024viw}, rotation of the particle plays a key role in solving the Humpty-Dumpty problem of rotational degrees of freedom. It is conceivable that the spin axis will wobble around the orientation of the magnetic field due to the precession and nutation of the particle. So, the spatial trajectories are affected, which can cause the Humpty-Dumpty effect in spatial degrees of freedom. Roughly considering, we can treat the wobbling of the spin axis as spin-dependent random noise to estimate the contrast loss due to the impact of the particle's rotation on the spatial motion. However, a separate dedicated study will be required. We will pursue these directions in separate publications.

%


\section*{ACKNOWLEDGMENTS}

A.M. would like to thank Sougato Bose for many helpful discussions on this research topic.
T.Z is supported by the China Scholarship Council (CSC). RR is supported by Beasiswa Indonesia Bangkit, the Ministry of Religious Affairs of the Republic of Indonesia (Kemenag), and the Indonesia Endowment Fund for Education (LPDP) of the Ministry of Finance of the Republic of Indonesia. AM's research is partly funded in part by the Gordon and Betty Moore Foundation through Grant GBMF12328, DOI 10.37807/GBMF12328. This material is based upon work supported by the Alfred P. Sloan Foundation under Grant No. G-2023-21130.
MS is supported by the National Research Foundation, Singapore through the National Quantum Office, hosted in A*STAR, under its center for Quantum Technologies Funding Initiative (S24Q2d0009); MS's research is supported by the Ministry of Education, Singapore under the Academic Research Fund Tier 1 (FY2022, A-8000988-00-00).


\appendix

\section{Quantum forced harmonic oscillator}\label{app:a}
Consider a harmonic oscillator perturbed by a time-dependent interaction from external sources; the Hamiltonian can be written as
\begin{equation}\label{ht}
    {H}(t)={H}_0+{V}(t) \, ,
\end{equation}
where the time-independent part ${H}_0$ is defined by
\begin{equation}\label{h0}
    {H}_0= \frac{{p}^2}{2m}+\frac{1}{2}m\omega^2{x}^2 \, .
\end{equation}
We here consider interactions linear in $x$, namely (see~\cite{lo1991generating} for more general case)
\begin{equation}\label{vt}
    {V}(t,x)= f(t){x}\, ,
\end{equation}
where the function $f(t)$ represents a time-varying external perturbation. In the interaction picture, denoted by the subscript "$I$", the time evolution operator ${U}_{I}(t)$ satisfies the equation
\begin{equation}\label{schrodingerequ}
    i\hbar \frac{\partial}{\partial t} {U}_I(t)= {H}_I (t) {U}_{I}(t)\, ,
\end{equation}
where
\begin{equation}\label{HIt}
    {H}_I(t)={U}_0^\dagger(t) {V}(t) {U}_0(t)\,, \quad {U}_0(t)=e^{-i{H}_0t/\hbar}\, .
\end{equation}
Note that generally, ${U}_I(t, t_0)$ and ${U}_0(t, t_0)$, but we have set $t_0 = 0$.
For the Hamiltonian given by Eqs.~\eqref{h0},~\eqref{vt} in the interaction picture (Eq.~\eqref{HIt}), the operator ${H}_I(t)$ can be solved as:
\begin{align}\label{Hi}
    {H}_I (t) &=f(t)\left[ {x}\cos(\omega t) + {p} \frac{\sin(\omega t)}{m\omega} \right]\nonumber\\
    &= \sqrt{\frac{\hbar}{2m\omega}}f(t)({a}\, e^{-i\omega t}+{a}^\dagger e^{i\omega t})\, ,
\end{align}
where we used the standard definition of the ladder operators in terms of the creation, ${a}^{\dagger}$, and annihilation, ${a}$, operators 
\begin{equation}
    {x} = \sqrt{\frac{\hbar}{2m\omega}}({a}+{a}^\dagger) \, ,\qq{} {p} = -i\sqrt{\frac{\hbar m\omega}{2}}({a}-{a}^\dagger)\, .
\end{equation}
From the Eqs.~\eqref{schrodingerequ},~\eqref{Hi}, it is easy to verify that the evolution operator ${U}_I$ is a displacement operator, which takes the form
\begin{equation}\label{Ui}
    {U}_I(t)=e^{i\varphi(t)}{D}(\zeta(t))\equiv e^{i\varphi(t)} {\rm exp}(\zeta {a}^\dagger-\zeta^* {a})\, ,
\end{equation}
where the complex number $\zeta(t)$ and the global phase $\varphi(t)$ obey the equations by solving Eqs.~(\ref{schrodingerequ}, \ref{Hi}, \ref{Ui}):
\begin{equation}\label{eq:zeta}
\left\{ 
\begin{aligned}
    &i\hbar \frac{\partial}{\partial t} \zeta(t)= \sqrt{\frac{\hbar}{2m\omega}}f(t) e^{i\omega t} \, , \\
    &-i\hbar \frac{\partial}{\partial t} \zeta^*(t)= \sqrt{\frac{\hbar}{2m\omega}}f(t) e^{-i\omega t} \, ,\\
    & \frac{\partial \varphi(t)}{\partial t}=\frac{i}{2}\left( \frac{\partial\zeta^*}{\partial t}\zeta - \frac{\partial\zeta}{\partial t}\zeta^*\right) \, , \\
    \end{aligned}
\right.
\end{equation}
where the solution of $\zeta$ and $\varphi$ can be solved by the integral equations:
\begin{align}\label{zetatheta}
    &\zeta(t)=-i\sqrt{\frac{1}{2\hbar m \omega}}\int_0^t dt'\, f(t')e^{i\omega t'}\, ,\nonumber\\
    &\varphi(t)=\frac{-1}{2\hbar m \omega}\int_0^t dt'\int_0^{t'}dt'' f(t')f(t''){\rm sin}(\omega(t'-t'')) \, ,
\end{align}
Now, the evolution operator in the Schr\"odinger picture is given by the total time evolution operator (combining the free evolution and interaction part):
\begin{equation}\label{evolutionoperator}
    {U}(t) = {U}_0(t){U}_I(t)= e^{i\varphi(t)} e^{-i\omega t {a}^\dagger {a}} {D}(\zeta(t)) \, .
\end{equation}
Considering an initial state coherent state $\lvert \alpha\rangle$, the evolution of the coherent state is given by 
\begin{align}\label{psit}
    \lvert\psi(t)\rangle &= {U}(t)\lvert\alpha\rangle = e^{i\varphi} e^{-i\omega t {a}^\dagger {a}} {D}(\zeta(t)){D}(\alpha)\lvert0\rangle \nonumber\\
    &= e^{i(\varphi +\varphi_\alpha)} e^{-i\omega t {a}^\dagger {a}} \lvert (\alpha+\zeta(t)) \rangle  \nonumber\\
    &= e^{i(\varphi +\varphi_\alpha)} \lvert (\alpha+\zeta(t)) e^{-i\omega t} \rangle \, ,
\end{align}
where $\varphi_\alpha$ denotes an additional phase
\begin{equation}
    \varphi_\alpha\equiv \frac{1}{2}( \zeta \alpha^* - \alpha \zeta^* )\, .
\end{equation}
Eq.~\eqref{psit} shows how the time-evolution displaces the initial state and is used in Eq.~\eqref{psiLpsiRSI} in the text.

\section{Transfer function }\label{app:b}

The spin-independent noise-induced phase difference $\delta\varphi$ is shown in Eq.~\eqref{deltaphiSI}. The statistical variance $\mathbb{E}[\delta\varphi^2]$ can be derived, see~\cite{Milburn:2015}
\begin{widetext}
\begin{align}
    \mathbb{E}[\delta\varphi^2]= & \ 4u^2\omega^4 \int_0^{\frac{2\pi}{\omega}} dt_1 dt_2 \int_0^{t_1} dt_1' \int_0^{t_2} dt_2' \mathbb{E}[[\Delta u(t_1)+\Delta u(t_1')][\Delta u(t_2)+\Delta u(t_2')]{\rm sin}\omega(t_1-t_1'){\rm sin}\omega(t_2-t_2')] \nonumber\\
    =& \ 4u^2\omega^4 \int_0^{\frac{2\pi}{\omega}} dt_1 dt_2 \int_0^{t_1} dt_1' \int_0^{t_2} dt_2' \left(\mathbb{E}[\Delta u(t_1)\Delta u(t_2)]+\mathbb{E}[\Delta u(t_1)\Delta u(t_2')] + \mathbb{E}[\Delta u(t_1')\Delta u(t_2)] + \mathbb{E}[\Delta u(t_1')\Delta u(t_2')]\right)  \nonumber\\
    &\times{\rm sin}\omega(t_1-t_1'){\rm sin}\omega(t_2-t_2') \nonumber\\
    =& \ 4u^2\omega^4 \int_0^{\frac{2\pi}{\omega}} dt_1 dt_2 \int_0^{t_1} dt_1' \int_0^{t_2} dt_2'\int_0^\infty\frac{d\Omega}{2\pi} S_{\Delta u}(\Omega)\left( e^{i\Omega (t_1-t_2)} + e^{i\Omega (t_1-t_2')} + e^{i\Omega (t_1'-t_2)} + e^{i\Omega (t_1'-t_2')}\right)  \nonumber\\
    & \times{\rm sin}\omega(t_1-t_1'){\rm sin}\omega(t_2-t_2') \nonumber\\
    =& \ \int_0^\infty d\Omega\, S_{\Delta u}(\Omega) \frac{32 u^2\omega^6}{\pi} \frac{{\rm sin}^2(\pi\Omega/\omega)}{(\Omega^3-\Omega\omega^2)^2} \, ,
\end{align}
\end{widetext}

Here, we have used the Wiener-Khinchin theorem (\eqref{PSD}), which states that the spectral density function $S_{\Delta u}(\Omega)$ is the Fourier transform of the autocorrelation function $\mathbb{E}[\Delta u(t) \Delta u(t')]$. Furthermore, we have used the idea that the integration over the four exponents can be combined due to the symmetry of their integration of $\Omega$.


Therefore, by definition of Eq.~\eqref{dephase}, the transfer function of the dephasing effect is given by
\begin{equation}
    F(\Omega) = \frac{32 u^2\omega^6}{\pi} \frac{{\rm sin}^2(\pi\Omega/\omega)}{(\Omega^3-\Omega\omega^2)^2} = \frac{32u^2}{\pi} \frac{{\rm sin}^2\left(\pi x \right)}{(x^3-x)^2}\, ,
\end{equation}
where $x=\Omega/\omega$.



\bibliography{reference}

\end{document}